\documentclass[preprint,journal]{vgtc}       

\ifpdf
  \pdfoutput=1\relax                   
  \usepackage{graphicx}                
  \DeclareGraphicsExtensions{.pdf,.png,.jpg,.jpeg} 
\else
  \usepackage{graphicx}                
  \DeclareGraphicsExtensions{.eps}     
\fi%

\graphicspath{{figures/}{pictures/}{images/}{./}} 

\usepackage{microtype}                 
\PassOptionsToPackage{warn}{textcomp}  
\usepackage{textcomp}                  
\usepackage{mathptmx}                  
\usepackage{times}                     
\usepackage{cite}                      
\usepackage{tabu}                      
\usepackage{booktabs}                  
\usepackage{enumitem}
\usepackage{epigraph} 
\usepackage[dvipsnames]{xcolor}
\usepackage[noorphans,vskip=0.5ex, font=itshape]{quoting} 
\usepackage{fontawesome}

\onlineid{0}

\vgtccategory{Research}
\vgtcpapertype{please specify}

\title{Visualization Design Practices in a Crisis: Behind the Scenes with COVID-19 Dashboard Creators}

\author{Yixuan Zhang, Yifan Sun, Joseph D. Gaggiano, Neha Kumar, Clio Andris, Andrea G. Parker}
\authorfooter{
\item
Yixuan Zhang, Joseph D. Gaggiano, Neha Kumar, Clio Andris, and Andrea G. Parker are with The Georgia Institute of Technology. E-mail: yixuan@gatech.edu; jgaggiano@gatech.edu; neha.kumar@gatech.edu; clio@gatech.edu; andrea@cc.gatech.edu.
\item
Yifan Sun is with The College of William \& Mary. E-mail: ysun25@wm.edu. |
}

\shortauthortitle{Visualization Design Practices in a Crisis \MakeLowercase{\textit{Zhang et al.}}: Visualization Design Practices in a Crisis: Behind the Scenes with COVID-19 Dashboard Creators}

\abstract{During the COVID-19 pandemic, a number of data visualizations were created to inform the public about the rapidly evolving crisis. Data dashboards, a form of information dissemination used during the pandemic, have facilitated this process by visualizing statistics regarding the number of COVID-19 cases over time. Prior work on COVID-19 visualizations has primarily focused on the design and evaluation of specific visualization systems from technology-centered perspectives. However, little is known about what occurs behind the scenes during the visualization creation processes, given the complex sociotechnical contexts in which they are embedded. Yet, such ecological knowledge is necessary to help characterize the nuances and trajectories of visualization design practices in the wild, as well as generate insights into how creators come to understand and approach visualization design on their own terms and for their own situated purposes. In this research, we conducted a qualitative interview study among dashboard creators from federal agencies, state health departments, mainstream news media outlets, and other organizations that created (often widely-used) COVID-19 dashboards to answer the following questions: how did visualization creators engage in COVID-19 dashboard design, and what tensions, conflicts, and challenges arose during this process? Our findings detail the trajectory of design practices---from creation to expansion, maintenance, and termination---that are shaped by the complex interplay between design goals, tools and technologies, labor, emerging crisis contexts, and public engagement. We particularly examined the tensions between designers and the general public involved in these processes. These conflicts, which often materialized due to a divergence between public demands and standing policies, centered around the type and amount of information to be visualized, how public perceptions shape and are shaped by visualization design, and the strategies utilized to deal with (potential) misinterpretations and misuse of visualizations. Our findings and lessons learned shed light on new ways of thinking in visualization design, focusing on the bundled activities that are invariably involved in human and nonhuman participation throughout the entire trajectory of design practice.}


\vgtcinsertpkg

\begin{document}


\firstsection{Introduction}

\maketitle

\epigraph{\textit{``Those who cannot remember the past are condemned to repeat it.''}}{--- George Santayana, Reason in Common Sense, 1905}

\begin{figure}[h]
  \centering
  \includegraphics[width=\linewidth]{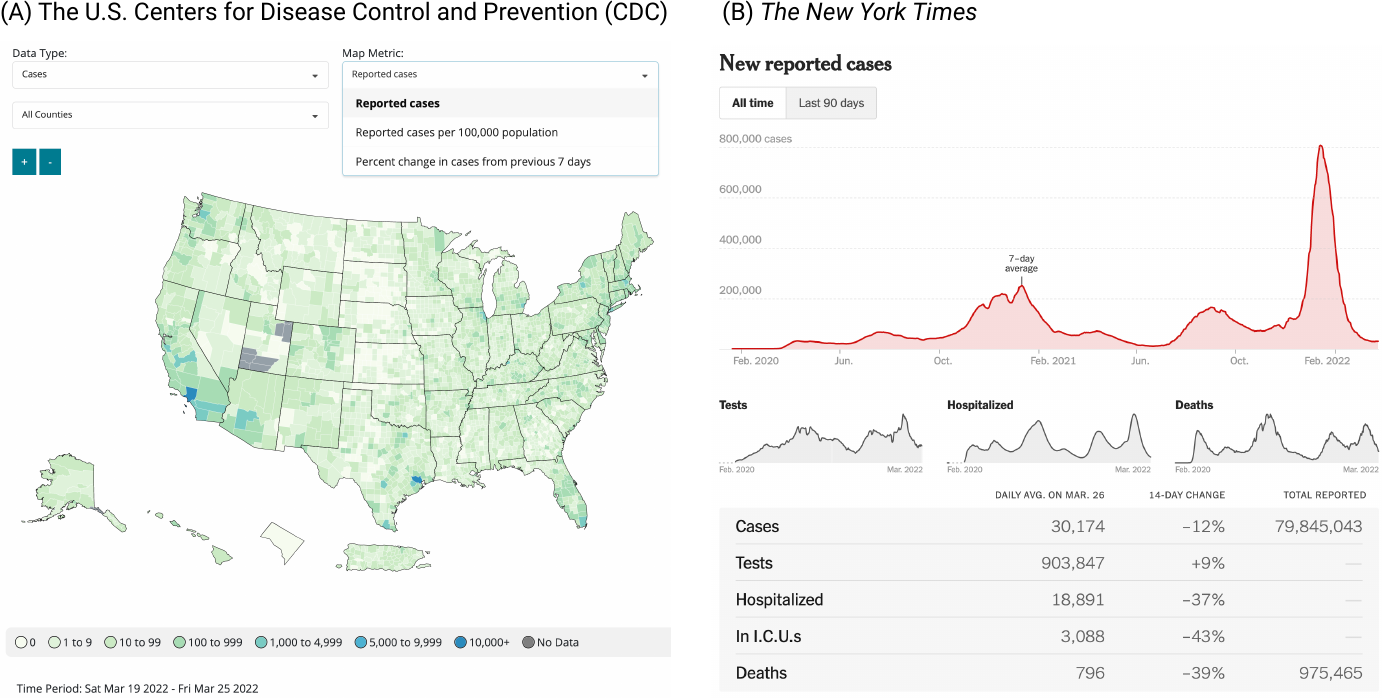}  
  \caption{Examples of COVID-19 dashboards displaying data on COVID-19 cases, deaths, hospitalizations, testing, and vaccinations take many forms such as geographical maps (A) and time-series charts (B).} 
  \label{fig:COVID_data_tracker} 
\end{figure}

\noindent
During the COVID-19 pandemic, \textit{COVID-19 dashboards}---a set of visualizations that display data on COVID-19 cases, deaths, hospitalizations, testing, and vaccinations~\cite{cdc_data_tracker}---were produced to help combat this public health crisis (see examples in \autoref{fig:COVID_data_tracker}). These visualizations play an essential role in facilitating top-down decision-making for policymakers and public health agencies, while simultaneously communicating the current pandemic situation to the general public to help inform their day-to-day decisions and guide their behaviors. Over the course of the pandemic, visualizations, such as COVID-19 dashboards, have become an integral component of crisis information infrastructures (i.e., networks of heterogeneous sociotechnical systems that facilitate information communication during crises)~\cite{zhao2022visualization}.

While prior work has proposed and evaluated specific designs of COVID-19 visualizations~\cite{bernasconi2021conceptual, boulos2020geographical, cay2020understanding, dixit2020rapid, fang2021evaluating, li2021visualizing}, little work has investigated the behind-the-scenes design practices employed by COVID-19 visualization creators. These practices include design activities undertaken to produce COVID-19 visualizations, as well as the struggles and conflicts that arose in their production---all of which are not fully visible when looking at the visualization artifacts themselves. Producing these COVID-19 visualizations was a massive collaborative activity, warranting the help of a wide range of stakeholders to contribute to the dashboard designs, ranging from visualization designers to leadership and epidemiologists~\cite{bowe2020learning}. The urgency of the pandemic situation and the complex collaboration necessary for creating these visualizations mean that the designers may not have fully followed the principles devised by visualization researchers. This reveals a unique opportunity to investigate how these crucial COVID-19 visualizations are designed and produced ``in the wild'' (e.g., how technology is used in the everyday/real world and in naturalistic settings~\cite{crabtree2013introduction, rogers2017research}). Given that other crises and pandemics, beyond COVID-19, will likely happen in the future~\cite{frutos2020covid}, learning from and summarizing past design practices and experiences is crucial to prepare for future crises. Such retrospective examination of design practices is critical to understanding how people approach and engage in visualization design practices in their everyday lives, on their own terms, and for their own situated purposes. 

This research seeks to examine the design practices involved in COVID-19 visualizations, focused on public-facing dashboards---one crucial form of information dissemination utilized during this pandemic---and to uncover the invisible forces behind these processes. This work is guided by the following research questions (RQs):
\newline
$\bullet$  RQ1. How did creators engage in the COVID-19 dashboard design?
\newline 
$\bullet$  RQ2. What challenges arose during the production of these visualizations in times of COVID-19?   

To answer these questions, we conducted a qualitative interview study with 26 participants who were involved in the production of public-facing COVID-19 dashboards. Participants were from organizations such as federal agencies, state health departments~\cite{states_health_departments}, mainstream news media, and other organizations that contributed to this core information infrastructure. Visualizations produced by our participants have all gained high visibility during this pandemic, ranging from thousands to hundreds of millions of daily visits. 

We contribute qualitative empirical research examining the visualization design practices during the COVID-19 public health crisis. Specifically, our findings detail the trajectory of visualization design practices through phases of creation, expansion, maintenance, and then termination, shaped by the evolving, complex interplay between design goals, visualization tools and technologies, labor, and public engagement. We examined the tensions between designers and the general public, the conflicts between shifting public demands and standing policies regarding the type and amount of information to be visualized, how visualizations shape and are shaped by public perception, and the strategies that designers use when dealing with (potential) misinterpretations and misuse of visualizations. Pointing out these bureaucratic tensions and limitations imposed on visualization design is essential as it can uncover shortcomings within the paradigms designated as ``best'' practices that may not appear in a lab setting. Consequently, practices need to be adapted within the existing infrastructure to keep up with rapidly evolving contexts. Our work also puts forth new ways to understand visualization design and raises questions about visualization design practices and ethical issues in design that depart from our habitual modes of design thinking.

\section{Background and Related Work}
\label{sec:related_work}
Below we discuss the role of visualizations in crisis communication and management, and research that examines visualization design practices. 

\subsection{Visualization in Crisis Communication \& Management}
\label{susbsec:vis_as_infrastructure}
 
Visualizations have played an increasingly critical role in crisis communication and management and have served as integral components of crisis information infrastructure~\cite{chen2020rampvis, soden2016infrastructure, zhao2022visualization}. Information infrastructures refer to networks of sociotechnical systems~\cite{lee2018bridge}. They are characterized by openness to users (no fixed notion of ``user'' in terms of types of and number of ``users''), interrelatedness of systems serving different purposes, and dynamically evolving systems shaped by existing systems and practices~\cite{monteiro2013artefacts}.  Information infrastructures used during crises are referred to as crisis information infrastructures~\cite{zhang2022shifting}. Visualization technologies help emergency response agencies communicate crisis and risk information to audiences~\cite{kwan2005emergency, konev2014run, padilla2017effects, waser2011nodes}; these technologies include real-time systems in the context of terrorist attacks~\cite{kwan2005emergency} and simulation-based visualization tools during prior pandemics~\cite{konev2014run, waser2011nodes}. Prior work has also explored data visualizations of historical pandemic events and other crises such as hurricanes~\cite{Bica2019Com, lu2004web, padilla2017effects, preim2020survey, welhausen2015visualizing}.

A growing body of work has explored the design and evaluation of COVID-19 data visualizations~\cite{boulos2020geographical, comba2020data, dixit2020rapid, fang2021evaluating, ivankovic2021features}. For example, a survey study~\cite{zhang2021mapping} curated and analyzed 668 COVID-19 data visualizations to map the landscape of existing visualizations. Their work summarized visualization techniques for communicating different messages, such as informing users about COVID-19 severity and forecasting trends. Other work has explored how visualizations impacted people's interpretations and risk perceptions regarding current pandemic situations~\cite{dixit2020rapid, li2021visualizing, romano2020covid, padilla2022impact}. For example, Li et al.~\cite{li2021visualizing} found that using contrasting colors in a sequential order for visualizing maps leads to high accuracy of interpretation among their participants. 
 
Despite prior work focusing on the visual design of COVID--19 visualizations, work that investigates the \emph{design practices} incorporated by the designers is sparse, given the complex sociotechnical contexts involved. This lack of research is compounded by the fact that limited documentation regarding COVID-19 visualization design process exists online, with notably infrequent blog posts publicized by teams working on the COVID-19 dashboards as some of the only documentation available~\cite{nyt_tracker_redesign}. This lack of knowledge in understanding the processes of producing these crucial visualizations restrains researcher's ability to learn from both the ``success'' and ``failure'' in the design of visualizations, as well as the challenges involved in these practices in the wild. Essentially, learning from history in a design context helps avoid repeated mistakes and mitigate risks for future crises~\cite{alkhatib2017examining, soden2021time, zhang2020understanding}.  
 
Additionally, while previous work has provided suggestions on how to design visualizations to help the general public understand information under non-crisis contexts~\cite{li2021visualizing, ma2011scientific, shanks2017teaching, van2014communicate}, little work has investigated the interplay between public perceptions and visualization design \emph{in the middle of a crisis}. Therefore, extending existing work focused on visualizations in crisis communication and addressing the above-mentioned gaps, our work seeks to examine the visualization design practices, considering the socio-cultural-political contexts in which they are embedded.

\subsection{Understanding Visualization Design Practices}
We situate our work in the existing research focused on design practices. Over the past decade, visualization scholars have examined design practices involved in creating visualizations~\cite{mckenna2014design, kang2011characterizing, parsons2020design, parsons2022understanding, sadowski2021anyway, tory2021finding, walny2019data}, shifting away from the focus on design studies: Visualization design studies are projects in which researchers analyze a specific real-world problem faced by domain experts, design a visualization system that supports solving this problem, validate the design, and reflect on lessons learned in order to refine visualization design guidelines~\cite{sedlmair2012design}, from either a technique-driven or interaction-centered perspective~\cite{kuutti2014turn, parsons2020design}. Compared to visualization design studies, work focused on design practices appears to be more \emph{distributed, embedded, and long-term}~\cite{kuutti2014turn, parsons2022understanding, stolterman2008nature}. For example, Kang et al.~\cite{kang2011characterizing} characterized the visual analytics process through a longitudinal field study, suggesting design implications backed by research to improve the process. Likewise, Tory et al.~\cite{tory2021finding} examined the work practices of dashboard users, reframing their practices as ``data conversations'' (i.e.  iterative interactions between people and data to ask and answer questions) to better address these users' diverse needs. Recent work by Parsons et al.~\cite{parsons2020design, parsons2022understanding} also explored how visualization practitioners engage with design, recognizing that design situations are complex and situated in real-world problems. Sadowski~\cite{sadowski2021anyway} argued that ``the dashboard is dead'', more specifically, that corporate dashboards are abandoned after due to interrelated organization-cultural-technical reasons, such as new leadership and issues on automated updates that cause existing dashboards to break and become irreparable.

Existing work examining design practices has also suggested that the paradigms known as ``best practices'' in academia are not always applicable in the wild. Challenges presented in real-world design practices illuminate important issues, such as a lack of logical methodology to guide the design process~\cite{parsons2022understanding}, issues involving tool switching when creating visualizations~\cite{kang2011characterizing, tory2021finding}, and insufficient support for collaboration~\cite{kang2011characterizing}. Furthermore, these studies collectively indicate that \emph{context matters} since knowledge and design are socially situated~\cite{bardzell2011towards, haraway1988situated, kuutti2014turn}. What people do in particular situations cannot only be attributed to standard procedures and plans~\cite{suchman1987plans, suhaimi2022investigating}. The context of the situation, such as contingencies, constraints, and problems that are only discovered and addressed once encountered in practice dictate design practices~\cite{dourish2004action}. 

So far, there has been a lack of research investigating the design practices in visualization production in times of the COVID-19 public health crisis, which is interwoven with the complex sociotechnical contexts. The burden of the pandemic on the design process should be addressed, as reflecting on this knowledge is essential to gain insights into the visualization production process in extensive and fast-evolving contexts and to reveal sociotechnical considerations for visualization design practices. Consequently, more realistic and effective methods can be identified through these processes. The lessons learned throughout the entire trajectory of COVID-19 dashboard production can also be valuable in combating future crises. Our research seeks to address these gaps.

\section{Method}
\label{sec:method} 
This study examines the behind-the-scenes values, directives, situations and imperatives that shaped the design of COVID-19 dashboards. We conducted a qualitative interview study with 26 participants in the United States, upon approval from our Institutional Review Board (IRB). Using a qualitative inquiry approach, we gained in-depth insights into participants' experiences, interpretations, attitudes, and values. 

\subsection{Participant Recruitment and Overview}
\label{subsec:participant_overview}
Our criteria for participation were that participants needed to have created public-facing visualizations that communicated COVID-19 data (i.e., COVID-19 dashboards) in the U.S.; be age 18 years or older; and be able to speak and write in English.

We used multiple methods to recruit participants. The initial participant pool was based on a recent survey paper of 668 public-facing COVID-19 visualizations~\cite{zhang2021mapping}, which included 158 U.S.-based distinct information outlets. We first reached out to all 158 design agencies who created COVID-19 visualizations via e-mail. In the e-mail, we provided descriptions of our research, a consent form, compensation information, a description of potential benefits, and an explanation of risks associated with participation. Additionally, in the e-mail sent to individuals who designed COVID-19 dashboards, we asked them to share our study information with others who might be interested in our study. This snowball sampling approach~\cite{goodman1961snowball} was particularly useful when a direct contact was not publicly available. We made a concerted effort to locate core design personnel during recruitment, especially those who worked in federal agencies and state health departments.  

In total, we virtually interviewed 26 participants between May and August 2021. Participants were from different organizations, including federal agencies that were involved in collecting, sharing, and producing COVID-19 related data products, state health departments~\cite{states_health_departments}, mainstream news media outlets, nonprofit organizations and research institutions, and emerging crowdsourced organizations that contributed to the federal agencies' COVID-19 data products. The COVID-19 dashboards produced by our participants have gained high visibility (e.g., up to hundreds of millions of website visits daily). Our participants also held different positions, ranging from the director of a state health department to a data journalist.
 
\autoref{tab:demographic} provides an aggregated overview of our participants' characteristics in terms of gender, age range, race, education level, organizations that participants worked for while producing COVID-19 dashboards, and amount of time spent working on the COVID-19 dashboards. On average, our participants had experiences in creating visualizations for 7.5 years with a standard deviation of 6 years. Participants held degrees in different domains, including epidemiology and/or public health (n=8), computer science (n=4), geography and/or city planning (n=4), design (n=4), cognitive science (n=2), math and statistics (n=2), economics (n=1), and religion (n=1). 
 
\begin{table}[tb] 
 \caption{Sample characteristics of interview participants.}
 \label{tab:demographic}
 \scriptsize%
	\centering%
 \begin{tabu}{%
	l%
	*{6}{l}%
	*{2}{r}%
	}
     \toprule
     \textbf{Participants'}    &  \textbf{Response Options}  &  \textbf{Number of} \\ 
     \textbf{Characteristics} &    &  \textbf{Participants}  \\
     \midrule
     \multicolumn{3}{c}{\textit{Demographics}} \\ [1.5ex] 
        Gender  & Women & 11 (42\%)  \\
                & Men	& 15  (58\%)  \\  
        \hline \\[-1ex] 
        Age Range	& 18-24	& 3 (12\%)  \\
                    & 25-34	& 16 (61\%)  \\
                    & 35-44	& 5 (19\%)  \\
                    & 45-54	& 1 (4\%)  \\
                    & 55+	& 1 (4\%)   \\
        \hline \\[-1ex] 
        Race & American Indian & 1 (4\%)  \\
             & Asian	       & 6 (23\%)  \\ 
             & Hispanic	       & 1 (4\%)  \\
             & White           & 18 (69\%) \\
        \hline \\[-1ex] 
        Education & Bachelor   & 9 (35\%)  \\
                  & Master 	   & 12 (46\%)  \\
                  & PhD 	   & 5 (19\%)  \\
     \midrule          
     \multicolumn{3}{c}{\textit{Experiences with COVID-19 dashboards}} \\ [1.5ex] 
        Organizations         & Federal \& state health departments     & 15 (58\%) \\
        that                  & News outlets                            & 4 (15\%) \\ 
        participants          & Crowdsourced organizations              & 3 (12\%) \\ 
        worked for            & Nonprofit \& research institutions      & 4 (15\%) \\ 
        \hline \\[-1ex] 
        Visualization             & Tableau                   & 13  (38\%) \\ 
        tools used*               & D3.js                     & 10  (29\%) \\ 
        by designers              & Microsoft Power BI        &  7  (21\%) \\  
                                  & ArcGIS                    &  3  (9\%) \\  
                                  & R Shiny                   &  1  (3\%) \\ 
        \multicolumn{3}{l}{* \small Note: In a few instances, participants used multiple tools.} \\ 
        \multicolumn{3}{l}{\small Therefore, the resulting total number shown in the last column was more than 26.} \\ 
        \hline \\[-1ex] 
        Duration working          & Less than 6 months       & 2  (8\%) \\
        on COVID-19               & 6-12 months              & 6  (23\%) \\ 
        dashboards                & More than 12 months      & 18 (69\%) \\ 
 \bottomrule
 \end{tabu} 
\end{table}

\subsection{Semi-structured Interview Study Procedure}
In preparation for the study, the research team developed a semi-structured interview guide to understand the design practices of COVID-19 dashboards. We customized the interview guide for each interview session, allowing us to dive deeper into the topic space with our participants. To develop each customized interview guide, the first author interacted with the COVID-19 dashboard(s), reading through the documentation and user guide (if any, such as related publications and online articles), and taking notes of notable questions to be asked during the interview. The structure of the interview includes 1) grounding the interview in the context of the dashboard(s) that the participant was involved in producing, 2) understanding the design practices, design decisions, and design strategies among participants, focused on an overview of the process, people involved, tools and technologies, and evaluation of COVID-19 visualizations, 3) examining issues and challenges specific to the COVID-19 crisis, and 4) general reflection and lessons learned. These interview topics were in line with prior work that examines design practices (as described in ~\autoref{sec:related_work}).
 
Prior to the interview, participants were asked to sign a consent form. Specific policies and regulations (e.g., certain federal agencies), restricted some participants from receiving compensation. Therefore, the consent form was presented in two separate ways: one with \$30 compensation and one without monetary compensation. 

All interviews were conducted remotely via Zoom (an online meeting tool). During the interview, participants were prompted to retrospectively reflect on the design practices in producing COVID-19 dashboards. Interview sessions lasted approximately 75 minutes, ranging between 60 minutes and 110 minutes, and were recorded. Participants were then asked to complete a short online demographic survey via Qualtrics~\cite{Qualtrics}. At the end of the study session, each eligible participant was compensated with a \$30 gift card. Participants who were restricted from receiving monetary compensation were thanked by the research team for their time and input. The semi-structured interview guide and demographic survey are provided in the supplementary material.

\subsection{Analysis}
All interview sessions (of a total 34 hours) were transcribed verbatim. To follow the anonymization policy, participants' names were replaced with codes, labeled as PID (e.g., P01, ..., P26). The General Inductive Approach~\cite{thomas2006general} was used to guide the thematic analysis of the transcripts. The lead author first read the transcripts closely to get an initial understanding of the data, and then labeled the segments of text of each page of the transcript to create categories, which were aggregated to develop low-level codes. Similar low-level codes were clustered together to achieve high-level themes. Throughout the analysis, the research team discussed and refined the emerging themes.

\subsection{Positionality Statement}
\label{subsec:reflexivity} 
All authors of this paper have experience conducting COVID-19 related research and hold different research backgrounds, ranging from data visualization, geographic information system, public health, health equity, social sciences, and crisis informatics. Our varied positions afford unique perspectives into the design practices surrounding COVID-19 dashboards, especially given the assorted sociotechnical factors present in the context of the COVID-19 public health crisis. The first author with interdisciplinary research backgrounds in visualization, crisis informatics, and health equity led the research by planning and conducting the research study; the rest constantly provided invaluable feedback on study design, data collection, and analysis.

\section{Findings} 
Our findings characterize visualization designers' shifting practices during the COVID-19 pandemic, from the creation of the dashboards (denoted as the creation phase), to the addition of new features that address the increasing demands from leadership and the general public (denoted as the expansion phase), to the reduction in the cost of maintaining and updating the dashboards (denoted as the maintenance phase), and finally to the conclusion in the development efforts (denoted as the termination phase). These practices were shaped by the change in the overarching goals, visualization tools and technologies, labor, public engagement, as well as the fast-evolving pandemic situations. We also present findings surrounding the conflicts and tensions between visualization designers and the general public that have emerged as the crisis develops. Furthermore, we describe participants' positionality regarding how participants' diverse affiliations may inherently lead to different stances and biases within their design practices.  

\begin{figure*}[h]
  \centering
  \includegraphics[width=\linewidth]{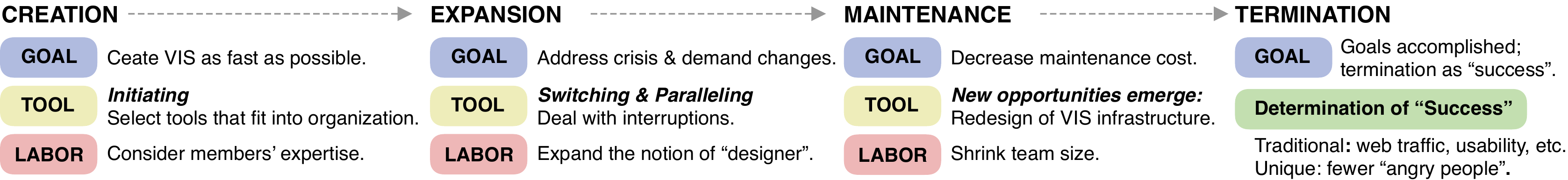}  
  \caption{A simplified illustration of phases involved in the design practices of COVID-19 visualization production.} 
  \label{fig:phases}
  \vspace{-10pt}
\end{figure*}

\subsection{Shifting Practices: Creation, Expansion, Maintenance, and Termination}   
Our findings suggest that design practices shifted over time in parallel with the fast-evolving pandemic situation, from creation, to expansion, maintenance, and termination of COVID-19 dashboards. 

\subsubsection{The Creation Phase}  
\label{subsubsec:creation}

\textbf{Goal:}
According to our participants, the overarching goal for their institutions was to create public-facing COVID-19 visualizations \emph{as fast as possible} in response to the emerging crisis and to keep the public informed as things were changing quickly during this pandemic.

\noindent
\textbf{Tool initiating:}  
We found that the choice of tools for creating COVID-19 dashboards for organizations was primarily driven by how well the tools fit into the organization's existing infrastructure that participants worked for to achieve their overall goal. For example, P09 explained why Microsoft Power BI was chosen,
\begin{quoting} 
    ``We use Power BI because we migrated to Office 365 last spring, and Power BI is built into that infrastructure. So suddenly, it was a datavis tool that everyone had access to.''
\end{quoting}
Many participants (n=19), like P09, mentioned accessibility as the most crucial factor determining how well the tool(s) fit into existing infrastructure. \textit{Accessibility} in this context refers to the ease of attainment of a technology for designers. The collaborative and distributed workflow involved in the creation of COVID-19 dashboards necessitates the need for technology to fit into the existing infrastructure. 

Apart from considerations regarding accessibility of visualization products and tools across team members, the level of expertise of team members also played a role in how these products fit into existing infrastructure. Given the time constraints felt during the pandemic, many designers indicated that they \textit{``did not want to learn new things [technologies] at this point''}. And yet, the selected tools were not necessarily the most preferred ones for the individual designers, according to our participants. The determination of visualization tool often came down to a compromise. Factors, such as the degree of mastery among core design teams and the level of top-down decisions and leadership involved in the design decisions, collectively contributed to the characterization of how well a tool fit into the infrastructure. 
 
\noindent
\textbf{Labor:}
At the very beginning of the creation phase, a core team was established primarily based on members' expertise and evaluation of available resources. The assessment of ``expertise'' varied depending on the nature of the organization. In our study, for public health departments and other organizations that had internal public health experts involved in the team, considerations of expertise include both prior experiences and knowledge in public health and technical skills in data analytics and visualization. For other types of organizations, the primary consideration was technical expertise.

\subsubsection{The Expansion Phase}
\label{subsubsec:expansion}
Soon after the initial dashboard creation, the number and type of designers and stakeholders, as well as the amount of attention from the general public, expanded at an unexpected rate. All participants indicated that the initial dashboards that were produced at an early phase became \textit{no longer sufficient} due to the fast-evolving nature of the pandemic. As more COVID-19 data came in alongside the increasing demands from organizational leadership and the general public, the number of visualization features and functions also increased tremendously. For example, features, such as the visualization of disaggregated demographic data (e.g., by age, gender, race and ethnicity) and the addition of dashboard components specifying the type of COVID-19 testing data, soon began to emerge. Consequently, as the crisis continued to change and the needs of the public increased, participants' workloads subsequently increased, forcing them to work for extended hours (e.g.,  60 hours above/ week). On top of this, the increasing demands also brought up a variety of ``unexpected'' interruptions (e.g., server jams, performance issues).  

The rapid growth in the scale of teams, the increased demand for more data and features, the growing attention yielded by these visualizations, and the rise in unexpected interruptions collectively mark the expansion phase. These changes in labor, demand from the public and leadership, together with the uncertain crisis also tremendously shaped what and how visualizations were produced.

\noindent
\textbf{Goal:} As such, a new overarching goal for organizations was to address the changing needs in visualizing more dimensions of COVID-19 data and improving visualization performance, shaped by the fast-evolving pandemic situation and demands from both leadership and the public. 

\noindent
\textbf{Tool switching and paralleling:}
While the initial driver to select specific visualization tools was primarily focused on how well the tools fit into an organization's existing infrastructure to meet the demands quickly, these demands shifted throughout the COVID-19 crisis. These shifts caused unexpected interruptions, as the existing solutions failed to address changing needs. Relatedly, tools and techniques used in the production of COVID-19 dashboards also changed. 

Our findings suggest that unexpected interruptions drove decisions to switch tools, such as server jams that caused failures in addressing the increasing public engagement. For example, a few participants (n=4) described situations in which their visualization systems crashed as they \textit{``did not expect that huge volume of users''}. As a result, the design team spent \textit{``a lot of time enlarging the database and server.''} 

Interruptions of this kind, in turn, prompted changes on the service provider side---companies who provided the infrastructure for creating visualizations, as P03 from a state health department recalled,
\begin{quoting}
    ``Over the course of this pandemic, we have switched from Tableau Public  
    to our own Tableau-facing servers because Tableau was complaining that we actually had some of the greatest volumes that they've ever had on to the Tableau Public, like literally they had to build out Tableau Public more just because of some of our dashboards.'' 
\end{quoting}

The increasing scale of COVID-19 dashboard projects also changed the fundamental goals for the visualizations. These goals shifted away from just \textit{``getting some dashboard out there''} to \textit{``caring about the performance''}. For example, some participants (n=3) told us that Tableau was their first choice by default. However, due to Tableau's performance issues, they had to duplicate the visualizations in D3, which allowed for better performance and usability, as P16 explained,
\begin{quoting} 
    ``The state dashboard has such a massive audience. And we had multiple visualizations from the same workbook that have made its performance really laggy and buggy honestly. We use Tableau for several months, while we were also building the D3 versions at the same time...''  
\end{quoting}
P16's experiences, in contrast with some other participants, illustrated another type of strategy in the development of COVID-19 dashboards: utilizing two different sets of tools and technologies in parallel. It is worth noting that this parallel work requires additional resources both in terms of labor and funding. 

\noindent
\textbf{Labor:}
During the expansion phase, the design teams of certain organizations with ample resources scaled up from a couple of people to hundreds of people. Newly involved designers also held different positions and experiences. Our participant (P11) described how designers expanded at an unexpected rate, influencing their design approaches,
\begin{quoting}
    ``When you have an infrastructure like [my organization], it's not just data visualization specialist creating the products; it's everybody creating them.'' 
\end{quoting} 
This increase in the scale and categories of team members reflects the change in the notion of ``designer''. Specifically, designers expanded from visualization specialists to a more broad sense of dashboard ``designers'' consisting of a variety of roles within an organization. These new ``designers'' were recruited as contractors and/or outsourced temporary staff; and some were digital volunteers who held other full-time daily jobs (e.g., as developers, journalists). Our observation is in line with our descriptions in the background and related work (\autoref{sec:related_work}) that states an information infrastructure is characterized by its openness in the definition of ``users'' (i.e., the number and types of ``users'' of visualization tools/ designers of visualizations). This shifting notion of designer also suggests that COVID-19 dashboards gained infrastructural properties over time. 

The expansion in labor involved in the creation of COVID-19 dashboards caused issues in visualization design, such as the inconsistency of visuals (e.g., marks and channels) even within the same organization. P11 explained,
\begin{quoting}
    ``A lot of those products [maps] are actually made by developers writing in JavaScript. Usually it's a not a GIS person; it's a development person that works with GIS data.''
\end{quoting}
This elaboration helped explain why certain choropleth maps did not use normalized data---a hotly debated topic in academia~\cite{Field_mappingblog} during the pandemic. As such, during the expansion phase, the shifting definition of visualization ``designer'' also changed how visualization artifacts were produced. 

\subsubsection{The Maintenance Phase}
\label{subsubsec:maintenance}
Later, we observed that COVID-19 visualization design practices shifted to a maintenance phase. This transition meant that the team no longer created new platforms but rather maintained what they had already built. The exact transition point that distinguishes the maintenance stage is challenging, if not impossible, to specify. Yet, participants indicated that 6 to 12 months after the first versions of the COVID-19 dashboards were released, they reached a plateau in terms of the labor and effort involved in this phase. We highlight that \emph{temporality} played a crucial role in this practice shift and was also reflected by the long-term nature of the pandemic and its enduring impacts.  

\noindent
\textbf{Goal:}
Design teams aimed to shrink the size of the team, decreasing the amount of effort and cost spent on the COVID-19 dashboards' maintenance work.

\noindent
\textbf{Labor:} During the maintenance phase, a few large-scale organizations (n= 3) decreased the total number of designers for producing COVID-19 visualizations. However, our participants indicated that the core team members remained the same. 

\noindent
\textbf{New technological opportunities emerge:}
During the maintenance phase, new and exciting opportunities for innovation emerged. New technologies and visualization techniques were adopted, such as a redesign of organization's visualization infrastructure and pipelines, an increase in the degree of automation (rather than manually checking data issues at the beginning of the pandemic), and acquisition of human-in-the-loop approaches. For example, P07 reflected, 
\begin{quoting}
    ``A lot of technical possibilities for us... like the pages will get republished multiple times per day, and it was really big laborious effort that's involved. Like 100 people [working] at various kinds [of work] so we're trying to scale that process down... So we've developed a new infrastructure over the past months that will make it easier to just keep publishing this page for longer periods with not many people having to spend so much time on it.''
\end{quoting} 
Participants also indicated that they had gradually gained foresight into potential future changes in data and technologies used for maintaining COVID-19 dashboards after the first couple of months experiencing constant changes in data format, data sources, and tool switches. Accordingly, participants described strategies to accommodate and act on potential future changes and thus, moved from a reactive to a proactive approach in managing uncertainty.

\noindent
\textbf{Breakdowns signaled the lapsing practices:}
However, participants also indicated that the maintenance phase also had unexpected issues and interruptions. While some interruptions remained isolated events and were quickly remedied (as mentioned in \autoref{subsubsec:expansion}), other breakdowns had cascade effects that fundamentally changed design workflows. Breakdowns occurred when design agencies were no longer able to access open-sourced tools and/or when the data sources relied upon stopped updating. For example, P14 told us, 
\begin{quoting}
    ``We had a free [visualization software] license due to COVID-19, [as] they were giving free access to some users, and they removed that access at some point... So we stopped using it to do those updates to [our] other dashboard.''
\end{quoting}
These findings shed light on the interdependence of infrastructure networks (e.g. the reliance of some organizations' COVID-19 dashboards on the operations of visualization tools and platforms) that complicated the maintenance work.

\subsubsection{The Termination Phase}
\label{subsubsec:termination}
Though only two participants in our study indicated that they stopped updating and maintaining their dashboards, a few more participants (n=3) pondered the question of ``when to stop''. This signals that the visualization design practice entered the termination phase. 

The termination of elements comprising the framework of organization's visualizations caused some breakdowns to occur, as some dashboards relied on publicly available data and open-source tools to function. However, decisions of termination were not necessarily fueled by completely negative circumstances. Our participants indicated that the cessation of design updates could mean that an organization's tasks and motivations were accomplished and organizations no longer felt the need to support their resource-demanding products. These perspectives emphasize the idea that the discontinuation of updates and maintenance may not always be a bad thing and should sometimes be viewed as a mark of ``success'', as P13 working in a crowdsourced team that contributes to the federal COVID-19 data products explained, 
\begin{quoting}
    ``We wanted to wind down and point [the audiences] to the federal government as they should be single source of truth for this kind of data...''  
\end{quoting}
P13 mentioned that one key driver of terminating their COVID-19 data products was because their team believed that \textit{``the data was good enough from the Federal Government that [they] don't have to [continue] doing this...''} These quotes also indicate a need to reduce the number of trustworthy sources presenting COVID-19 data in the U.S. context. After all, the diverse sources may lead to confusion and even distrust in authorities~\cite{zhang2022shifting}. Yet, while the termination of dashboard updates was considered as ``success'' for some groups, others suffered with this change. This conflicting perspective in the metrics of ``success'' sheds light on the need to expand our considerations when evaluating ``success'', as different communities may have different evaluation standards toward the same metric. 

\noindent
\textbf{Determination of ``success'':}
Participants mentioned that the determination of success among COVID-19 visualizations included traditional metrics, such as the number of web traffic/views, total times of media coverage, number of ``embeds'' (if applicable), usability (e.g., visualizations work properly throughout the course of the pandemic), and clarity (e.g., less clarification questions). These metrics were similar to those used in academia~\cite{shamim2015evaluation, ridley20208}. 

However, we also found one ``unique'' consideration in determination the ``success'' of visualizations---\emph{fewer ``angry people''}. For example, P04 who worked in state public health department explained to us,  
\begin{quoting}
    ``The measure of successes was a lack of angry people... Unfortunately, a reasonable volume of really nasty comments came in from the public because [they felt] we weren't giving them exactly what they wanted...''  
\end{quoting} 

We argue that this metric of fewer ``angry people'' was not simply a proxy to ``user satisfaction'' (defined as how well users liked using the system to complete visualization tasks~\cite{shamim2015evaluation}). Instead, this measure of ``a lack of angry people'' not only signifies a shift in focus to the emotional aspects of assessing visualizations but also sheds light on the ongoing struggles with external stressors from the general public. This sense of struggle also reflects the tensions seen within the relationships between visualization designers and the general public (which will be unpacked in depth in \autoref{subsec:conflicts_designers_public}).    

\vspace{5pt}
\noindent
\textbf{Summary:} 
We have presented findings about the shifting practices from visualization creation to expansion, maintenance, and termination, as well as the inherent challenges and opportunities unveiled within these processes. Findings in this subsection mostly focus on visualization producers. In the next subsection, we broaden the focus to the relationship between the producers and the public to examine the how visualizations shape and, simultaneously, are shaped by public engagement with visualizations. 

\subsection{Conflicts between VIS Designers and General Public}  
\label{subsec:conflicts_designers_public}
Tensions and conflicting perspectives exist between designers and the general public within visualization design practices and are further complicated by the existing policies and regulations, as well as the unprecedented uncertainty and stress, that the COVID-19 crisis has engendered. Examining these conflicts reveal that visualizations in times of COVID-19 are the result of negotiations between designers, the general public/audiences, and standing policies. 

\subsubsection{Conflicts between Public Demands and Standing Policies}  
Our findings show that conflicts arose between public demands and standing policies regarding COVID-19 dashboard design. On the one hand, the general public requested more data and transparency, specifically lesser degrees of granularity data~\footnote{Data granularity in healthcare is defined as
``the level of detail at which the attributes and characteristics of data quality in healthcare data are defined''~\cite{davoudi2015data}.}, from the design agencies and organizations. On the other hand, the designers indicated that they could not fulfill these public requests given their existing privacy rules and regulations. This conflict was particularly prevalent among federal agencies and state health departments.  

\noindent
\textbf{The interplay between visualization design, policy making, and public demands:}
When participants reflected upon the evolution of COVID-19 dashboards, almost all of them indicated that their dashboards slowly became more granular and close-to-real-time. But the practices of producing visualizations in low-level data granularity were also largely constrained by the standing policies and procedures. For example, P03 working in a state public health department said,
\begin{quoting} 
    ``When we began this [dashboard], it was very much based on standing policy and procedure. We obviously want to preserve the anonymity of individuals that have been reported to us and minimize any likelihood that they could inadvertently or subsequently be identified. Historically we have not wanted to provide `real time counts' down to anything. We really progressed from a very high level to a very fine grain pretty quickly, which required a lot of changes to how we approached the business we typically do.''
\end{quoting} 

While the general public urgently requested more data to become available for review, government agencies simultaneously encountered uncertainty regarding how much information they could show to the public; a situation in which P03 described as a \textit{``fog of war''} to accurately represent the state of ambiguity these agencies found themselves in. Specifically, P03 described the organizational deliberation that occurred as they considered whether or not to visualize zip code-level data,
\begin{quoting}
    ``People were demanding data, down to the zip code level, and you owned a lot of that transition to that level of displaying that information... we have unfortunately kind of the fog of war that's going on...'' 
\end{quoting}
Yet, participants also mentioned that they needed to follow methods for de-identification of protected health information in accordance with the Health Insurance Portability and Accountability Act (HIPAA) Privacy Rule~\footnote{The Health Insurance Portability and Accountability Act of 1996 (HIPAA) is a federal law that required the creation of national standards to protect sensitive patient health information from being disclosed without patients' consent~\cite{HPAA}.}. For example, P22 explained to us, 
\begin{quoting}
    ``Usually we go by a rule of any cell less than six [items/people], where it goes like if it's by county/ sex/ age/ etc... once we start breaking/stratifying down, anything that's less than six, we don't share because it becomes there's a risk of identity occasion. And how we got to that number is based on some HIPAA rules, which note that there can't be a higher risk of identification than 4\%.''
\end{quoting}

However, with the increasing public request of lower-level granularity of data and associated dashboards, existing policies began to change and adapt to address this demand. P03 continued,
\begin{quoting}
    ``As we evolved to reporting cases by zip code that really there was the need, first and foremost, to influence leadership to make this change, and then we had to develop a decision memo for our Commissioner to sign off and approve that he authorized the release of data at this level of granularity which was essentially did not align with policy, so we basically had to get all that foundational paperwork in place. Now we're obviously concurrently building the tools at the same time, to report on it, so that, when it's approved we're ready to roll but it certainly was not a `hey that sounds like a cool idea, let's do that digitally' as it went down to that in granularity...''
\end{quoting}
As P03 described, rather than mere changes to data processing and visualizations, this process involved invisible work, such as negotiating and conversing with leadership and preparing paperwork, all while simultaneously working on newer versions of the dashboards, such as visualizing lower-granularity COVID-19 data. Over time, COVID-19 visualizations have become \emph{boundary objects} that support conversations between stakeholders who hold divergent viewpoints. Star and Griesemer~\cite{star1989institutional} describe ``boundary object'' as artifacts that are used for practices not necessarily agreed upon by all the people who use them but to bridge knowledge across different stakeholders and facilitate shared understanding and collaboration surrounding a common goal. We can conceptualize these visualizations as boundary objects in times of crisis, as they are the results of a continuous negotiation between designers, the general public or the audience, and the standing policies and regulations. 

\noindent
\textbf{Discrepancies in policy interpretations led to diverse practices:}
However, the design practices and decisions surrounding the granularity of data visualizations differed across agencies and organizations. According to our participants, for example, the decisions made by design agencies that were fully HIPAA-covered were more constrained than those who were not fully HIPAA-covered. In other words, some designers had different constraints due to the HIPAA rules, which affected the degree to which information could be detailed when visualizing the data. For example, P13 told us,
\begin{quoting} 
    ``We actually had some opportunity to look at what other states are doing and when we were trying to first develop our dashboards, [a particular state (name omitted)] was publishing the line-level data with their dashboard, so one line per case and it included things like if that person died, what their date of death was. And I remember looking at this and just being like oh my god, who made this decision, how did they come to this ability to publish data at that granular level? We would never do that ourselves...''
\end{quoting}
The lack of cohesive understanding and precision in the existing rules and policies led to the design of a diverse set of COVID-19 dashboards. Yet, these discrepancies may cause more severe impacts, such as stimulating distrust in authorities who dealt with COVID-19 data products, as recent work suggests~\cite{zhang2022shifting}. 

\subsubsection{Public Perceptions Shape \& are Shaped by Visualizations}   
\label{subsubsec:public_reactions_shape_design}
The challenges and conflicts associated with the production of COVID-19 dashboards eventually shifted from technological considerations regarding what and how to visualize data to considerations regarding the emotions and feelings of people. Our participants' experiences with direct and indirect interactions with audiences demonstrated the ways in which visualization designers are influenced by people's emotional reactions and how their design practices shifted over time as they considered addressing the negative feelings of the public. 

Our participants described situations in which their design decisions were directly shaped by public perceptions and emotions during COVID-19. For example, P14 from a state health department described, 
\begin{quoting}
    ``Very early on like January 2021, there was a lot of anxiety about where the  [COVID-19] vaccine was going, [vaccine] availability and [people were asking] `is it in my neighborhood? how many doses? why did my county [have less] but that county got that many?' which there are inherent flaws [regarding] how the data dashboards are provided... In some instances, very large [vaccine] doses got sent to health systems that had a centralized hub, which in turn distributed out to other localities but it made it look bad. 
    
    At one point [a county name (omitted)] had more doses than anywhere else in the state, and it was just purely because one of the health systems was up there. And we ultimately got asked to take down that component of a dashboard for the distribution because showing vaccine doses by locality was causing enough anxiety in some groups. They [leadership] said just pull it down, and there was some kickback of from that. Because folks [general public] were upset, and they felt that they couldn't see what was happening, but it was also painting a picture that wasn't truly reflective of what was happening in the community. So that was one of those where you have to take the backlash from the public about why, why, why can't I have this, why can't I see it. But it was better to not show that because it really wasn't telling the right story.''
\end{quoting}
These quotes shed light on the misalignment between data and people's interpretations inherent in visualizations. Design considerations taken into account when developing features for the COVID-19 dashboards were the result of people's reactions and emotions, shifting away from technological considerations. The main objective in creating these visualizations was to produce the ``right'' story that aims to help the audience understand the pandemic situation. Of course, the definition of ``right'' stories varies depending on context and is also relationally dependent on whose perspectives we examine from. And yet, as our findings suggest, the (potential) emotions and feelings aroused by visualizations amongst the general public are important to consider, especially given the extreme uncertainty and fast evolution of the ongoing crisis.

Public perceptions and emotions not only shaped the ways in which dashboards were iterated upon but also influenced the design decisions around future data products. For example, P23 shared a lesson learned regarding how information provided to the general public should be accompanied by proper context.  
\begin{quoting}
    ``I think lessons learned are how we need to change things: there were moments where there were bad things that happened like a teacher calling, saying, `I'm scared to go to my school because there's a lot of black kids. The teacher was saying that data shows that black people are more likely to get COVID'. And I was like, `No, that's not what wanted. We don't want people to be feared. So that's like, okay, we need to have some data product or weekly summary interpreting what that actually means. And so we did a whole series of summary [to explain] why these communities see more disparities and more COVID cases.''  
\end{quoting}
In line with our participants' descriptions, Black and Hispanic communities in the U.S. experienced disproportionately high rates of COVID-19-related deaths~\cite{reyes2020disproportional}. P23 describes how these health disparities, which were salient in COVID-19 dashboards, led to problematic reactions such as blanket fear of racial groups that experienced a higher burden of COVID-19 infection and death. These disparities during COVID-19 reflect longstanding inequalities in health, which are created by social and structural determinants of health~\cite{braveman2011social}. 

Simply compiling data and visualizing ``as is'' could stigmatize certain communities, leading people to stereotype, negatively view, look down upon, or discriminate against others. These emotional effects reflect a major limitation of data dashboards: a loss of context. To address a lack of context within COVID-19 dashboards, supplementary data products that provide additional details, such as the weekly summaries that our participants mentioned, have emerged over time. These supplementary products aimed to show how a variety of factors, such as reduced healthcare access and structural inequalities, could contribute to health inequities across socioeconomic, racial, and ethnic groups during the pandemic. These findings also echo calls-for-action to address health disparities~\cite{lopez2021racial}.  

Our findings show how data not only can inform or misinform but also shape beliefs, attitudes, and fears that result in some communities being further stigmatized and marginalized. As such, a visualization should not be conceptualized merely as a neutral object but instead, should be considered a tool that may engender emotions, feelings, and values among its audience. The emotions and feelings evoked by visualizations not only influence the design practices but also impact the design decisions of future data products.

\subsubsection{Dealing with Misinterpretations \& Misuse of Visualizations}
As previously described, participants mentioned that some features were intentionally avoided (e.g., mapping low-level granularity of vaccine availability data), as the resulting visualizations did not communicate the ``right'' message and invoked fear amongst the general public. This finding is also related to another emerging theme: the expectations of potential misinterpretations and misuse of data and visualizations during COVID-19. 

Participants, particularly those who had a background in epidemiology and public health, highlighted the importance of understanding the nature of COVID-19 data. For example, P22 said, 
\begin{quoting}
    ``The benefit of being an epidemiologist as well [was] a better understanding of the foundation of the data, the materials, more than anything else because there's a point at which it's not just data and visualization; there needs to be the awareness and understanding of what the data themselves represent in order to best represent it to the world.''
\end{quoting} 
Because these visualizations were produced by a wide range of individuals, authorities, and organizations from different backgrounds, there may have been a lack of sufficient understanding of COVID-19 data. P22 continued explaining her concern,
\begin{quoting} 
    ``There are a lot of armchair epidemiologists that don't necessarily have the background in the type of test matters as to how we define a case and the time between when you were exposed and when you develop symptoms matters, [and] all of those kinds of questions that we're used to dealing with. So we wanted to provide a little bit more context... We can't stop them from trying to introduce themselves, so at least we can provide as much guidance as we can.'' 
\end{quoting}
During the COVID-19 pandemic, the term \textit{``armchair epidemiologists''} has been used to describe anyone who has taken an interest in understanding the vast amounts of publicly-available open-sourced COVID-19 data~\cite{armchair2020Weinman}. Most emerging armchair epidemiologists are often ``legitimate experts'' in other fields, such as technologists, economists, and doctors who specialize in different fields of medicine; with some being ``COVID Influencers''~\cite{armchair2020Weinman}. To mitigate the risks of misinterpretation and misuse of COVID-19 data and visualizations, participants working within public health sectors sought to provide more contextual information about the dashboards, such as adding clear definitions of each term being used, and providing their methodology for classifying a COVID-19 ``case''.

\vspace{5pt}
\noindent
\textbf{Summary:}
Our findings suggest that the evolution of visualization design practices during COVID-19 involved tensions between the general public and designers. First, we observed the conflicts between public demands that requested an increasing amount and granularity of data, while standing policies and rules constrained designers in providing and visualizing this data. And yet, the interpretation of standing policies varied across organizations, leading to different design practices and the resulting visualization artifacts. Regardless of the continuation of tensions, public demands also prompted revisions made to existing policies and rules. Finally, our findings show that public perceptions shaped and were shaped by visualizations, demonstrating that visualization design practices shifted away from technological considerations to broader social-cultural aspects, taking into account audiences' emotions and lived experiences. These findings also indicate that visualizations during COVID-19, conceptualized as ``boundary objects'' are the results of a continuous negotiation between human actors (e.g., designers, the general public) and non-human actors (e.g., standing policies).  

\subsection{Participants' Positionality Regarding Design Practices}  
Given that much of the knowledge and lessons learned and shared by our participants are subjective, there is a need to reflect on how participants' professional backgrounds, their affiliations, and values may have impacted the production of COVID-19 visualizations. Below we describe some noticeable patterns within participants' positionality in relation to their design practices.  

\noindent
\textbf{Turn to humanity and sympathy:} 
Regardless of the organization that they worked for, almost all participants indicated that they were \textit{``extra careful and respectful''} during the dashboard production process. Participants indicated that it was important to \textit{``remember that the number is a person who's gotten sick and died''}, rather than just viewing the data as \textit{``printing numbers.''} This notion of humanity and sympathy should be a priority for designers creating visualizations, particularly during a pandemic. 

\noindent
\textbf{Are visualizations political and biased?}
One may believe that visualizations are inherently political and biased, as Winner would argue~\cite{winner1980artifacts}. In other words, one argument is that visualizations are (consciously or unconsciously) designed and deployed to favor certain social interests particularly within the politically polarized COVID-19 pandemic context~\cite{kerr2021political}. Yet, our participants indicated that their dashboard design process was \textit{``not a political thing''}, as this process aligned with their overall ``neutral'' design philosophy. Participants affiliated with governments mentioned additional pressure while working on the COVID-19 dashboards: \textit{``in contemporary American context, government isn't necessarily viewed in the most positive of light.''} Tensions, such as the public criticisms, misinterpretations, and feelings of distrust, made some of our participants feel \textit{``frustrated or exhausted''}. Likewise, regardless of the public perception about media bias and general media distrust during COVID-19~\cite{van2021public}, three out of four participants who worked in mainstream media mentioned that they tried to communicate the COVID-19 data through \textit{``simple and bare-bone''} visualizations. Although all of our participants did not indicate strong and intended narratives, they might have overlooked certain underlying or accompanying messages behind the visualizations during the design processes. For example, our finding suggest simply plotting data ``as is'' led to further stigmatization and discrimination (see \autoref{subsubsec:public_reactions_shape_design}). Audiences might simply take the face value of the message communicated through visualizations and consider it as ``data fact'' or the ``truth''. And yet, such takeaways might not reflect the truth in real life.
 
\noindent
\textbf{Different local constraints and enablers: } 
However, there are noticeable differences across participants' practices. Design practices are shaped by resources such as funding, human labor and expertise in public health and visualization technologies, as well as the level of bureaucratic tensions and constraints imposed on their design practices. These factors are both local ``constraints'' and ``enablers'' in defining design practices. For example, participants affiliated with government agencies and state health departments needed to follow HIPPA policies and Americans with Disabilities Act (ADA) regulations~\cite{cook1991americans} as compared to other participants. And yet, the extent to which designers followed these rules and the interpretations in how these rules are applicable in producing COVID-19 visualizations varied. However, these rules also enable responsive and accessible visualizations that can benefit broader audiences (e.g., those with color vision deficiencies).  

\section{Discussion}
\label{sec:discussion} 
Building upon our findings, we further reflect on the lessons learned and suggest opportunities for studying visualization design practices at large as well as addressing ethical issues involved in this process.  

\subsection{Shifting Attention to Visualization Design Practices}
Prior work suggests that a paradigm shift has occurred in visualization research~\cite{parsons2022understanding}, where the first wave focused on architectural models, the second wave focused on design studies, and the third wave, or ``practice paradigm''~\cite{kuutti2014turn}, is seeking to understand design practices in which tools, technical systems, organizational structures, and social and political environments are inherently interrelated and interwoven~\cite{goodman2011understanding, kuutti2014turn}. However, this shifting scope and lens into design practices have not been widely acknowledged within the VIS community~\cite{parsons2022understanding}. Our work attempts to address this gap and contributes to the examination of visualization design practices in the wild during COVID-19. 

When considering design practice as a unit of analysis, we no longer put the design artifact at front and center, while treating everything else as context~\cite{kuutti2014turn}. Instead, examining design practices in the wild necessitates attention to be focused on the bundled activities that are invariably involved in human and nonhuman participation, incorporating the social, cultural, and political contexts in which these activities are situated. Examining design practices also means identifying the \emph{relationship} that is mutually constitutive across actors, tools/technologies, and situations, as well as examining interrelated limitations and constraints. Below we re-iterate on the concepts illustrated through our findings and discuss their implications for future work. 

\textbf{Shifting phases.}
Temporality plays an essential role in studying design practices, and hence, it is important to examine the notion of ``shifts'', as our findings highlighted. These shifts in overarching design goals and tasks, tools and technologies, labor, socio-cultural-political environments, collectively suggest the need to adapt existing visualization practices in order to form new practices that better cope with the shifts. Turning our attention to the shifting practices also means approaching visualization design practices \textit{as they may become}---envisioning new designing practices that may be different from the current practices may help us mitigate the risk of future crises. Meanwhile, shifting practices also demand new forms of visualization research to uncover the challenges and opportunities throughout the life cycle of visualization design practices during crises.

\textbf{Shifting notion of ``designers''.} 
Throughout the COVID-19 dashboard design processes, the type and number of ``designers'' shifted rapidly and unexpectedly. This expansion of designers (e.g., the inclusion of people from diverse backgrounds) led to different practices utilized and design inconsistencies (e.g., color schemes, choice of maps). The discrepancy in design practices may have severe impacts on public perception and belief, generating a sense of distrust towards design agencies~\cite{zhang2022shifting}. Therefore, as an initial step to addressing the challenges arising from the shifting notion of ``designers'', work needs to examine how design tools support onboarding activities in order to effectively help newcomers get started with design, particularly under time pressure and stress. A successful onboarding is vital to ensure more cohesive design practices. Additionally, future work needs to consider how design is affected, shaped, and constrained by this shifting notion of designers. Addressing this matter will allow for more feasible and robust design practices in the future, particularly in situations of urgency and within teams experiencing high membership fluidity.   

\textbf{Shifting and paralleling tools.}
Switching tools for visualization is not an uncommon practice, as prior work has also suggested~\cite{tory2021finding}. However, developing (almost) the same visualizations using different tools in parallel might be less familiar. We consider simultaneous visualization development for the same overarching design goal as a risk mitigation strategy, ensuring high performance but making it functional first. This finding implies that for service providers (e.g., Tableau, MS Power BI, etc.), performance and accessibility issues need to continue to be addressed within their products. Additionally, designers and organizational leadership need to be aware of the long-term limitations of the tools they are using, taking into account the possibility that parallel work may be necessary, particularly in dynamic and uncertain contexts, and eventually turning to ``sustainable'' visualization practices. 

\textbf{Shifting dynamics between the public, designers, and policies.}
Our findings suggest that COVID-19 visualizations are becoming boundary objects as results of continuous negotiations between human actors/stakeholders who may hold divergent viewpoints and non-human actors such as existing policies and regulations. Seeing visualizations as boundary objects provides a new way of thinking about visualization design---how people from different fields with different expertise can bridge their separate knowledge domains, create a shared understanding, and improve decision making and policymaking. Viewing visualizations in times of crisis as boundary objects also indicates that ``plasticity''~\cite{star1989institutional} and flexibility become important properties of visualizations. Simply put, visualizations during crises should be ``plastic'' enough to adapt to local needs and the constraints of different stakeholders, yet robust enough to achieve a shared overarching goal.

\subsection{Towards Responsible Visualization Practices} 
Ethical issues in design have gained an increasing amount of attention within and outside of the VIS community~\cite{correll2019ethical, Correll_medium, Makulec_medium}. Our work contributes to this body of knowledge, presenting empirical findings that unpack the invisible forces and ethical issues behind the visualization design practices during COVID-19. Below, we discuss some considerations in approaching responsible visualization design practices. 

\textbf{Visualization is about data awareness.}
In the context of our research, data awareness refers to the deep understanding of what the data truly represents---the meaning of data. This data awareness goes beyond data type and data abstraction~\cite{munzner2014visualization} that visualization researchers are already familiar with. Instead, it exists in a particular \emph{context}. To obtain this data awareness, our participants highlighted the importance of leveraging domain knowledge in public health, geographic information systems, and so forth. A lack of domain knowledge, particularly during a public health crisis, increases the risk of (potential) misinterpretation and misuse of visualizations. As such, we argue that although visualizations \emph{can} theoretically be made for just about anything, this does not mean that they \emph{should} be made. Reflecting upon the consequences of creating visualizations is vital. One way visualization researchers can incorporate this moving forward is by reflecting upon positionality in future work both in academia and in the wild. Doing so will not only help the designers gain awareness of how their (implicit) biases may influence their design practices but will also encourage transparency regarding the limitations of design work to the audiences. 

\textbf{Addressing misinterpretation and misuse as design goals.}
Relatedly, with the prevalence of visualizations, misinterpretations and misuse of existing visualizations inevitably occur~\cite{doan2021misrepresenting}. Though it is challenging to solve these misinterpretations and misuse consequences fundamentally, some measures may help mitigate these risks, such as providing contextual information and methodology to the audiences. We suggest that future work considers addressing misinterpretation and misuse of visualizations as design goals. 

\textbf{Towards (truly) addressing inequity through visualization.}
We argue that extra care must be taken when producing visualizations related to health disparities, discrimination, and racial gaps, as ``unintended consequences'' can be detrimental to specific populations. During COVID-19, many agencies visualized COVID-19 data to understand the impact of the pandemic, presumably for ``social good''. However, our findings show how visualizations might have inadvertently facilitated further marginalization of specific communities over others. These ``unintended consequences'' may have negatively contributed to more serious societal issues at the root of design decisions. In-depth reflection on the potential repercussions of visualizations needs to be performed by designers and complete clarity within the visualizations should be a top priority among visualization designers. We should avoid ``superficially'' plotting the data ``as is'' but instead dig into ``how they may become'' and why they become what they are, with and through visualizations. Accordingly, one open question that we urge future work to examine is how visualizations can be designed and communicated to truly address inequity, if possible. Addressing this issue is a crucial step in examining the societal impact of visualizations.

\section{Limitations}
Our research is limited by the sampling methods we utilized, as we focused on design agencies in the United States. We encourage future research to examine the design practices of other visualizations types (e.g., visualizations that simulate infection and model parameters, visualizations that were used for internal purposes), as well as larger sample sizes and designers from non-western countries. Additionally, given that temporality plays an important role in understanding design practices, future work should continue to investigate the termination phase of practices when the pandemic is over, and the long-term effects of policy changes on visualization design in practice. Future work may also examine the relationship between the termination of visualization design practices and the existence of a variety of competitors, and how being in different roles in the design team influences their engagement with data-related policy-making processes. 

\section{Conclusion} 
Our work contributes to early research examining the design practices in the production of COVID-19 visualizations---a crucial component of crisis information infrastructure, considering the broader sociotechnical issues involved in these processes. The lessons learned from this work can be used to inform visualization design in future emergent situations and evolving contexts. While a large portion of design culture often focuses just on the creation of new visualizations, we encourage visualization communities to adequately capture the design practices involved in not only the creation phase, but also the expansion, maintenance, and termination phases. The entangled relationship between tools and technologies, labor, overarching goals, and public engagement shifted alongside specific contexts in which these activities were situated and thus collectively shaped the design practices. Finally, we call for future research to examine the sociotechnical perspective when studying the entire trajectory of visualization design practices in the wild, as well address ethical issues involved in these processes. 

\acknowledgments{
This material is based on work that is partially funded by an unrestricted gift from Google. We thank our participants for their help with our research. We also thank Georgia Tech Visualization Lab, the Wellness Technology Lab, and Jun Yuan from New York University for feedback; Carl DiSalvo for suggestions on this project; as well as our reviewers for their reviews. 
} 

\bibliographystyle{abbrv-doi}

\bibliography{ref}

\begin{thebibliography}{10}

\bibitem{alkhatib2017examining}
A.~Alkhatib, M.~S. Bernstein, and M.~Levi.
\newblock Examining crowd work and gig work through the historical lens of
  piecework.
\newblock In {\em Proceedings of the 2017 CHI Conference on Human Factors in
  Computing Systems}, CHI '17, p. 4599–4616. Association for Computing
  Machinery, New York, NY, USA, 2017. doi: {{%
10\hspace{.1pt}\discretionary{.}{%
}{.}\hspace{.4pt}1145\discretionary{/}{%
}{/}3025453\hspace{.1pt}\discretionary{.}{%
}{.}\hspace{.4pt}3025974}}


\bibitem{nyt_tracker_redesign}
W.~Andrews.
\newblock Why we redesigned the virus trackers.
\newblock https://www.nytimes.com/2021/04/01/us/covid-tracker-redesign.html,
  2022.

\bibitem{bardzell2011towards}
S.~Bardzell and J.~Bardzell.
\newblock Towards a feminist hci methodology: Social science, feminism, and
  hci.
\newblock In {\em Proceedings of the SIGCHI Conference on Human Factors in
  Computing Systems}, CHI '11, p. 675–684. Association for Computing
  Machinery, New York, NY, USA, 2011. doi: {{%
10\hspace{.1pt}\discretionary{.}{%
}{.}\hspace{.4pt}1145\discretionary{/}{%
}{/}1978942\hspace{.1pt}\discretionary{.}{%
}{.}\hspace{.4pt}1979041}}


\bibitem{bernasconi2021conceptual}
A.~Bernasconi and S.~Grandi.
\newblock A conceptual model for geo-online exploratory data visualization: The
  case of the covid-19 pandemic.
\newblock {\em Information}, 12(2):69, 2021. doi: {{%
10\hspace{.1pt}\discretionary{.}{%
}{.}\hspace{.4pt}3390\discretionary{/}{%
}{/}info12020069}}


\bibitem{Bica2019Com}
M.~Bica, J.~L. Demuth, J.~E. Dykes, and L.~Palen.
\newblock Communicating hurricane risks: Multi-method examination of risk
  imagery diffusion.
\newblock In {\em Proceedings of the 2019 CHI Conference on Human Factors in
  Computing Systems}, CHI '19, p. 1–13. Association for Computing Machinery,
  New York, NY, USA, 2019. doi: {{%
10\hspace{.1pt}\discretionary{.}{%
}{.}\hspace{.4pt}1145\discretionary{/}{%
}{/}3290605\hspace{.1pt}\discretionary{.}{%
}{.}\hspace{.4pt}3300545}}


\bibitem{boulos2020geographical}
M.~N.~K. Boulos and E.~M. Geraghty.
\newblock Geographical tracking and mapping of coronavirus disease
  covid-19/severe acute respiratory syndrome coronavirus 2 (sars-cov-2)
  epidemic and associated events around the world: how 21st century gis
  technologies are supporting the global fight against outbreaks and epidemics,
  2020. doi: {{%
10\hspace{.1pt}\discretionary{.}{%
}{.}\hspace{.4pt}1186\discretionary{/}{%
}{/}s12942\discretionary{%
}{-}{-}020\discretionary{%
}{-}{-}00202\discretionary{%
}{-}{-}8}}


\bibitem{bowe2020learning}
E.~Bowe, E.~Simmons, and S.~Mattern.
\newblock Learning from lines: Critical covid data visualizations and the
  quarantine quotidian.
\newblock {\em Big data \& society}, 7(2):2053951720939236, 2020. doi: {{%
10\hspace{.1pt}\discretionary{.}{%
}{.}\hspace{.4pt}1177\discretionary{/}{%
}{/}2053951720939236}}


\bibitem{braveman2011social}
P.~Braveman, S.~Egerter, and D.~R. Williams.
\newblock The social determinants of health: coming of age.
\newblock {\em Annual review of public health}, 32:381--398, 2011.

\bibitem{cay2020understanding}
D.~Cay, T.~Nagel, and A.~E. Yanta{\c{c}}.
\newblock Understanding user experience of covid-19 maps through remote
  elicitation interviews, 2020.

\bibitem{cdc_data_tracker}
{Centers for Disease Control and Prevention}.
\newblock Covid-19 data tracker.
\newblock https://covid.cdc.gov/covid-data-tracker, 2022.

\bibitem{HPAA}
{Centers for Disease Control and Prevention}.
\newblock Health insurance portability and accountability act of 1996 (hipaa).
\newblock https://www.cdc.gov/phlp/publications/topic/hipaa.html, 2022.

\bibitem{chen2020rampvis}
M.~Chen, A.~Abdul-Rahman, D.~Archambault, J.~Dykes, A.~Slingsby, P.~D. Ritsos,
  T.~Torsney-Weir, C.~Turkay, B.~Bach, A.~Brett, et~al.
\newblock Rampvis: Towards a new methodology for developing visualisation
  capabilities for large-scale emergency responses.
\newblock {\em arXiv preprint arXiv:2012.04757}, 2020.

\bibitem{comba2020data}
J.~L.~D. Comba.
\newblock Data visualization for the understanding of covid-19.
\newblock {\em Computing in Science \& Engineering}, 22(6):81--86, 2020. doi:
  {{%
10\hspace{.1pt}\discretionary{.}{%
}{.}\hspace{.4pt}1109\discretionary{/}{%
}{/}MCSE\hspace{.1pt}\discretionary{.}{%
}{.}\hspace{.4pt}2020\hspace{.1pt}\discretionary{.}{%
}{.}\hspace{.4pt}3019834}}


\bibitem{cook1991americans}
T.~M. Cook.
\newblock The americans with disabilities act: The move to integration.
\newblock {\em Temp. LR}, 64:393, 1991.

\bibitem{Correll_medium}
M.~Correll.
\newblock Visualization design principles for the pandemic.

\bibitem{correll2019ethical}
M.~Correll.
\newblock Ethical dimensions of visualization research.
\newblock In {\em Proceedings of the 2019 CHI Conference on Human Factors in
  Computing Systems}, CHI '19, p. 1–13. Association for Computing Machinery,
  New York, NY, USA, 2019. doi: {{%
10\hspace{.1pt}\discretionary{.}{%
}{.}\hspace{.4pt}1145\discretionary{/}{%
}{/}3290605\hspace{.1pt}\discretionary{.}{%
}{.}\hspace{.4pt}3300418}}


\bibitem{crabtree2013introduction}
A.~Crabtree, A.~Chamberlain, R.~E. Grinter, M.~Jones, T.~Rodden, and Y.~Rogers.
\newblock Introduction to the special issue of “the turn to the wild”,
  2013. doi: {{%
10\hspace{.1pt}\discretionary{.}{%
}{.}\hspace{.4pt}1145\discretionary{/}{%
}{/}2491500\hspace{.1pt}\discretionary{.}{%
}{.}\hspace{.4pt}2491501}}


\bibitem{davoudi2015data}
S.~Davoudi, J.~A. Dooling, B.~Glondys, T.~D. Jones, L.~Kadlec, S.~M. Overgaard,
  K.~Ruben, and A.~Wendicke.
\newblock Data quality management model (2015 update)-retired.
\newblock {\em Journal of AHIMA}, 86(10):expanded--web, 2015.

\bibitem{dixit2020rapid}
R.~A. Dixit, S.~Hurst, K.~T. Adams, C.~Boxley, K.~Lysen-Hendershot, S.~S.
  Bennett, E.~Booker, and R.~M. Ratwani.
\newblock Rapid development of visualization dashboards to enhance situation
  awareness of covid-19 telehealth initiatives at a multihospital healthcare
  system.
\newblock {\em Journal of the American Medical Informatics Association},
  27(9):1456--1461, 2020. doi: {{%
10\hspace{.1pt}\discretionary{.}{%
}{.}\hspace{.4pt}1093\discretionary{/}{%
}{/}jamia\discretionary{/}{%
}{/}ocaa161}}


\bibitem{doan2021misrepresenting}
S.~Doan.
\newblock Misrepresenting covid-19: lying with charts during the second golden
  age of data design.
\newblock {\em Journal of Business and Technical Communication}, 35(1):73--79,
  2021. doi: {{%
10\hspace{.1pt}\discretionary{.}{%
}{.}\hspace{.4pt}1177\discretionary{/}{%
}{/}1050651920958392}}


\bibitem{dourish2004action}
P.~Dourish.
\newblock {\em Where the action is: the foundations of embodied interaction}.
\newblock MIT press, 2004.

\bibitem{fang2021evaluating}
H.~Fang, S.~Xin, H.~Pang, F.~Xu, Y.~Gui, Y.~Sun, and N.~Yang.
\newblock Evaluating the effectiveness and efficiency of risk communication for
  maps depicting the hazard of covid-19.
\newblock {\em Transactions in GIS}, 2021. doi: {{%
10\hspace{.1pt}\discretionary{.}{%
}{.}\hspace{.4pt}1111\discretionary{/}{%
}{/}tgis\hspace{.1pt}\discretionary{.}{%
}{.}\hspace{.4pt}12814}}


\bibitem{Field_mappingblog}
K.~Field.
\newblock Mapping coronavirus, responsibly.
\newblock
  https://www.esri.com/arcgis-blog/products/product/mapping/mapping-coronavirus-responsibly,
  2020.

\bibitem{frutos2020covid}
R.~Frutos, M.~Lopez~Roig, J.~Serra-Cobo, and C.~A. Devaux.
\newblock Covid-19: the conjunction of events leading to the coronavirus
  pandemic and lessons to learn for future threats.
\newblock {\em Frontiers in medicine}, 7:223, 2020. doi: {{%
10\hspace{.1pt}\discretionary{.}{%
}{.}\hspace{.4pt}3389\discretionary{/}{%
}{/}fmed\hspace{.1pt}\discretionary{.}{%
}{.}\hspace{.4pt}2020\hspace{.1pt}\discretionary{.}{%
}{.}\hspace{.4pt}00223}}


\bibitem{goodman2011understanding}
E.~Goodman, E.~Stolterman, and R.~Wakkary.
\newblock Understanding interaction design practices.
\newblock In {\em Proceedings of the SIGCHI Conference on Human Factors in
  Computing Systems}, pp. 1061--1070, 2011. doi: {{%
10\hspace{.1pt}\discretionary{.}{%
}{.}\hspace{.4pt}1145\discretionary{/}{%
}{/}1978942\hspace{.1pt}\discretionary{.}{%
}{.}\hspace{.4pt}1979100}}


\bibitem{goodman1961snowball}
L.~A. Goodman.
\newblock Snowball sampling.
\newblock {\em The annals of mathematical statistics}, pp. 148--170, 1961.

\bibitem{haraway1988situated}
D.~Haraway.
\newblock Situated knowledges: The science question in feminism and the
  privilege of partial perspective.
\newblock {\em Feminist studies}, 14(3):575--599, 1988. doi: {{%
10\hspace{.1pt}\discretionary{.}{%
}{.}\hspace{.4pt}2307\discretionary{/}{%
}{/}3178066}}


\bibitem{ivankovic2021features}
D.~Ivankovi{\'c}, E.~Barbazza, V.~Bos, {\'O}.~B. Fernandes, K.~J. Gilmore,
  T.~Jansen, P.~Kara, N.~Larrain, S.~Lu, B.~Meza-Torres, et~al.
\newblock Features constituting actionable covid-19 dashboards: descriptive
  assessment and expert appraisal of 158 public web-based covid-19 dashboards.
\newblock {\em Journal of medical Internet research}, 23(2):e25682, 2021. doi:
  {{%
10\hspace{.1pt}\discretionary{.}{%
}{.}\hspace{.4pt}2196\discretionary{/}{%
}{/}25682}}


\bibitem{kang2011characterizing}
Y.-a. Kang and J.~Stasko.
\newblock Characterizing the intelligence analysis process: Informing visual
  analytics design through a longitudinal field study.
\newblock In {\em 2011 IEEE Conference on Visual Analytics Science and
  Technology (VAST)}, pp. 21--30, 2011. doi: {{%
10\hspace{.1pt}\discretionary{.}{%
}{.}\hspace{.4pt}1109\discretionary{/}{%
}{/}VAST\hspace{.1pt}\discretionary{.}{%
}{.}\hspace{.4pt}2011\hspace{.1pt}\discretionary{.}{%
}{.}\hspace{.4pt}6102438}}


\bibitem{kerr2021political}
J.~Kerr, C.~Panagopoulos, and S.~van~der Linden.
\newblock Political polarization on covid-19 pandemic response in the united
  states.
\newblock {\em Personality and Individual Differences}, 179:110892, 2021.

\bibitem{konev2014run}
A.~Konev, J.~Waser, B.~Sadransky, D.~Cornel, R.~A. Perdigão, Z.~Horváth, and
  M.~E. Gröller.
\newblock Run watchers: Automatic simulation-based decision support in flood
  management.
\newblock {\em IEEE Transactions on Visualization and Computer Graphics},
  20(12):1873--1882, 2014. doi: {{%
10\hspace{.1pt}\discretionary{.}{%
}{.}\hspace{.4pt}1109\discretionary{/}{%
}{/}TVCG\hspace{.1pt}\discretionary{.}{%
}{.}\hspace{.4pt}2014\hspace{.1pt}\discretionary{.}{%
}{.}\hspace{.4pt}2346930}}


\bibitem{kuutti2014turn}
K.~Kuutti and L.~J. Bannon.
\newblock The turn to practice in hci: Towards a research agenda.
\newblock In {\em Proceedings of the SIGCHI Conference on Human Factors in
  Computing Systems}, CHI '14, p. 3543–3552. Association for Computing
  Machinery, New York, NY, USA, 2014. doi: {{%
10\hspace{.1pt}\discretionary{.}{%
}{.}\hspace{.4pt}1145\discretionary{/}{%
}{/}2556288\hspace{.1pt}\discretionary{.}{%
}{.}\hspace{.4pt}2557111}}


\bibitem{kwan2005emergency}
M.-P. Kwan and J.~Lee.
\newblock Emergency response after 9/11: the potential of real-time 3d gis for
  quick emergency response in micro-spatial environments.
\newblock {\em Computers, Environment and Urban Systems}, 29(2):93--113, 2005.
  doi: {{%
j\hspace{.1pt}\discretionary{.}{%
}{.}\hspace{.4pt}compenvurbsys\hspace{.1pt}\discretionary{.}{%
}{.}\hspace{.4pt}2003\hspace{.1pt}\discretionary{.}{%
}{.}\hspace{.4pt}08\hspace{.1pt}\discretionary{.}{%
}{.}\hspace{.4pt}002}}


\bibitem{lee2018bridge}
C.~P. Lee and K.~Schmidt.
\newblock A bridge too far?: Critical remarks on the concept of
  ``infrastructure'' in computer-supported cooperative work and information
  systems.
\newblock In {\em Socio-Informatics}. Oxford University Press, 2018. doi: {{%
10\hspace{.1pt}\discretionary{.}{%
}{.}\hspace{.4pt}1093\discretionary{/}{%
}{/}oso\discretionary{/}{%
}{/}9780198733249\hspace{.1pt}\discretionary{.}{%
}{.}\hspace{.4pt}003\hspace{.1pt}\discretionary{.}{%
}{.}\hspace{.4pt}0006}}


\bibitem{li2021visualizing}
R.~Li.
\newblock Visualizing covid-19 information for public: Designs, effectiveness,
  and preference of thematic maps.
\newblock {\em Human Behavior and Emerging Technologies}, 3(1):97--106, 2021.
  doi: {{%
10\hspace{.1pt}\discretionary{.}{%
}{.}\hspace{.4pt}1002\discretionary{/}{%
}{/}hbe2\hspace{.1pt}\discretionary{.}{%
}{.}\hspace{.4pt}248}}


\bibitem{lopez2021racial}
L.~Lopez, L.~H. Hart, and M.~H. Katz.
\newblock Racial and ethnic health disparities related to covid-19.
\newblock {\em Jama}, 325(8):719--720, 2021. doi: {{%
10\hspace{.1pt}\discretionary{.}{%
}{.}\hspace{.4pt}1001\discretionary{/}{%
}{/}jama\hspace{.1pt}\discretionary{.}{%
}{.}\hspace{.4pt}2020\hspace{.1pt}\discretionary{.}{%
}{.}\hspace{.4pt}26443}}


\bibitem{lu2004web}
X.~Lu.
\newblock Web-gis-based sars epidemic situation visualization.
\newblock In {\em Fourth International Conference on Virtual Reality and Its
  Applications in Industry}, vol. 5444, pp. 445--452. International Society for
  Optics and Photonics, SPIE, 1000 20TH ST, Bellingham, WA 98225-6705, 2004.
  doi: {{%
10\hspace{.1pt}\discretionary{.}{%
}{.}\hspace{.4pt}1117\discretionary{/}{%
}{/}12\hspace{.1pt}\discretionary{.}{%
}{.}\hspace{.4pt}561185}}


\bibitem{ma2011scientific}
K.-L. Ma, I.~Liao, J.~Frazier, H.~Hauser, and H.-N. Kostis.
\newblock Scientific storytelling using visualization.
\newblock {\em IEEE Computer Graphics and Applications}, 32(1):12--19, 2012.
  doi: {{%
10\hspace{.1pt}\discretionary{.}{%
}{.}\hspace{.4pt}1109\discretionary{/}{%
}{/}MCG\hspace{.1pt}\discretionary{.}{%
}{.}\hspace{.4pt}2012\hspace{.1pt}\discretionary{.}{%
}{.}\hspace{.4pt}24}}


\bibitem{Makulec_medium}
A.~Makulec.
\newblock Ten considerations before you create another chart about covid-19.
\newblock
  https://medium.com/nightingale/ten-considerations-before-you-create-another-chart-about-covid-19-27d3bd691be8,
  2020.

\bibitem{mckenna2014design}
S.~McKenna, D.~Mazur, J.~Agutter, and M.~Meyer.
\newblock Design activity framework for visualization design.
\newblock {\em IEEE Transactions on Visualization and Computer Graphics},
  20(12):2191--2200, 2014. doi: {{%
10\hspace{.1pt}\discretionary{.}{%
}{.}\hspace{.4pt}1109\discretionary{/}{%
}{/}TVCG\hspace{.1pt}\discretionary{.}{%
}{.}\hspace{.4pt}2014\hspace{.1pt}\discretionary{.}{%
}{.}\hspace{.4pt}2346331}}


\bibitem{monteiro2013artefacts}
E.~Monteiro, N.~Pollock, O.~Hanseth, and R.~Williams.
\newblock From artefacts to infrastructures.
\newblock {\em Computer supported cooperative work (CSCW)}, 22(4):575--607,
  2013. doi: {{%
10\hspace{.1pt}\discretionary{.}{%
}{.}\hspace{.4pt}1007\discretionary{/}{%
}{/}s10606\discretionary{%
}{-}{-}012\discretionary{%
}{-}{-}9167\discretionary{%
}{-}{-}1}}


\bibitem{munzner2014visualization}
T.~Munzner.
\newblock {\em Visualization analysis and design}.
\newblock CRC press, 2014.

\bibitem{padilla2022impact}
L.~Padilla, H.~Hosseinpour, R.~Fygenson, J.~Howell, R.~Chunara, and E.~Bertini.
\newblock Impact of covid-19 forecast visualizations on pandemic risk
  perceptions.
\newblock {\em Scientific reports}, 12(1):1--14, 2022. doi: {{%
10\hspace{.1pt}\discretionary{.}{%
}{.}\hspace{.4pt}1038\discretionary{/}{%
}{/}s41598\discretionary{%
}{-}{-}022\discretionary{%
}{-}{-}05353\discretionary{%
}{-}{-}1}}


\bibitem{padilla2017effects}
L.~M. Padilla, I.~T. Ruginski, and S.~H. Creem-Regehr.
\newblock Effects of ensemble and summary displays on interpretations of
  geospatial uncertainty data.
\newblock {\em Cognitive research: principles and implications}, 2(1):1--16,
  2017. doi: {{%
10\hspace{.1pt}\discretionary{.}{%
}{.}\hspace{.4pt}1186\discretionary{/}{%
}{/}s41235\discretionary{%
}{-}{-}017\discretionary{%
}{-}{-}0076\discretionary{%
}{-}{-}1}}


\bibitem{parsons2022understanding}
P.~Parsons.
\newblock Understanding data visualization design practice.
\newblock {\em IEEE Transactions on Visualization and Computer Graphics},
  28(1):665--675, 2022. doi: {{%
10\hspace{.1pt}\discretionary{.}{%
}{.}\hspace{.4pt}1109\discretionary{/}{%
}{/}TVCG\hspace{.1pt}\discretionary{.}{%
}{.}\hspace{.4pt}2021\hspace{.1pt}\discretionary{.}{%
}{.}\hspace{.4pt}3114959}}


\bibitem{parsons2020design}
P.~Parsons, C.~M. Gray, A.~Baigelenov, and I.~Carr.
\newblock Design judgment in data visualization practice.
\newblock In {\em 2020 IEEE Visualization Conference (VIS)}, pp. 176--180,
  2020. doi: {{%
10\hspace{.1pt}\discretionary{.}{%
}{.}\hspace{.4pt}1109\discretionary{/}{%
}{/}VIS47514\hspace{.1pt}\discretionary{.}{%
}{.}\hspace{.4pt}2020\hspace{.1pt}\discretionary{.}{%
}{.}\hspace{.4pt}00042}}


\bibitem{preim2020survey}
B.~Preim and K.~Lawonn.
\newblock A survey of visual analytics for public health.
\newblock {\em Computer Graphics Forum}, 39(1):543--580, 2020. doi: {{%
10\hspace{.1pt}\discretionary{.}{%
}{.}\hspace{.4pt}1111\discretionary{/}{%
}{/}cgf\hspace{.1pt}\discretionary{.}{%
}{.}\hspace{.4pt}13891}}


\bibitem{Qualtrics}
{Qualtrics}.
\newblock Survey software: The best tool \& platform.
\newblock https://www.qualtrics.com/core-xm/survey-software/.

\bibitem{reyes2020disproportional}
M.~V. Reyes.
\newblock The disproportional impact of covid-19 on african americans.
\newblock {\em Health and Human Rights}, 22(2):299, 2020.

\bibitem{ridley20208}
A.~L. Ridley and C.~Birchall.
\newblock 8. evaluating data visualization: Broadening the measurements of
  success.
\newblock {\em Data Visualization in Society}, 127, 2020.

\bibitem{rogers2017research}
Y.~Rogers and P.~Marshall.
\newblock Research in the wild.
\newblock {\em Synthesis Lectures on Human-Centered Informatics}, 10(3):i--97,
  2017. doi: {{%
10\hspace{.1pt}\discretionary{.}{%
}{.}\hspace{.4pt}2200\discretionary{/}{%
}{/}S00764ED1V01Y201703HCI037}}


\bibitem{romano2020covid}
A.~Romano, C.~Sotis, G.~Dominioni, and S.~Guidi.
\newblock The scale of covid-19 graphs affects understanding, attitudes, and
  policy preferences.
\newblock {\em Health Economics}, 29(11):1482--1494, 2020. doi: {{%
10\hspace{.1pt}\discretionary{.}{%
}{.}\hspace{.4pt}1002\discretionary{/}{%
}{/}hec\hspace{.1pt}\discretionary{.}{%
}{.}\hspace{.4pt}4143}}


\bibitem{sadowski2021anyway}
J.~Sadowski.
\newblock ‘anyway, the dashboard is dead’: On trying to build urban
  informatics.
\newblock {\em New Media \& Society}, p. 14614448211058455, 2021. doi: {{%
10\hspace{.1pt}\discretionary{.}{%
}{.}\hspace{.4pt}1177\discretionary{/}{%
}{/}14614448211058455}}


\bibitem{sedlmair2012design}
M.~Sedlmair, M.~Meyer, and T.~Munzner.
\newblock Design study methodology: Reflections from the trenches and the
  stacks.
\newblock {\em IEEE Transactions on Visualization and Computer Graphics},
  18(12):2431--2440, 2012. doi: {{%
10\hspace{.1pt}\discretionary{.}{%
}{.}\hspace{.4pt}1109\discretionary{/}{%
}{/}TVCG\hspace{.1pt}\discretionary{.}{%
}{.}\hspace{.4pt}2012\hspace{.1pt}\discretionary{.}{%
}{.}\hspace{.4pt}213}}


\bibitem{shamim2015evaluation}
A.~Shamim, V.~Balakrishnan, and M.~Tahir.
\newblock Evaluation of opinion visualization techniques.
\newblock {\em Information visualization}, 14(4):339--358, 2015. doi: {{%
10\hspace{.1pt}\discretionary{.}{%
}{.}\hspace{.4pt}1177\discretionary{/}{%
}{/}1473871614550537}}


\bibitem{shanks2017teaching}
J.~D. Shanks, B.~Izumi, C.~Sun, A.~Martin, and C.~Byker~Shanks.
\newblock Teaching undergraduate students to visualize and communicate public
  health data with infographics.
\newblock {\em Frontiers in public health}, 5:315, 2017. doi: {{%
10\hspace{.1pt}\discretionary{.}{%
}{.}\hspace{.4pt}3389\discretionary{/}{%
}{/}fpubh\hspace{.1pt}\discretionary{.}{%
}{.}\hspace{.4pt}2017\hspace{.1pt}\discretionary{.}{%
}{.}\hspace{.4pt}00315}}


\bibitem{soden2016infrastructure}
R.~Soden and L.~Palen.
\newblock Infrastructure in the wild: What mapping in post-earthquake nepal
  reveals about infrastructural emergence.
\newblock In {\em Proceedings of the 2016 CHI Conference on Human Factors in
  Computing Systems}, CHI '16, p. 2796–2807. Association for Computing
  Machinery, New York, NY, USA, 2016. doi: {{%
10\hspace{.1pt}\discretionary{.}{%
}{.}\hspace{.4pt}1145\discretionary{/}{%
}{/}2858036\hspace{.1pt}\discretionary{.}{%
}{.}\hspace{.4pt}2858545}}


\bibitem{soden2021time}
R.~Soden, D.~Ribes, S.~Avle, and W.~Sutherland.
\newblock Time for historicism in cscw: An invitation.
\newblock {\em Proc. ACM Hum.-Comput. Interact.}, 5(CSCW2), oct 2021. doi: {{%
10\hspace{.1pt}\discretionary{.}{%
}{.}\hspace{.4pt}1145\discretionary{/}{%
}{/}3479603}}


\bibitem{star1989institutional}
S.~L. Star and J.~R. Griesemer.
\newblock Institutional ecology,translations' and boundary objects: Amateurs
  and professionals in berkeley's museum of vertebrate zoology, 1907-39.
\newblock {\em Social studies of science}, 19(3):387--420, 1989. doi: {{%
10\hspace{.1pt}\discretionary{.}{%
}{.}\hspace{.4pt}1177\discretionary{/}{%
}{/}030631289019003001}}


\bibitem{stolterman2008nature}
E.~Stolterman.
\newblock The nature of design practice and implications for interaction design
  research.
\newblock {\em International Journal of Design}, 2(1), 2008.

\bibitem{suchman1987plans}
L.~A. Suchman.
\newblock {\em Plans and situated actions: The problem of human-machine
  communication}.
\newblock Cambridge university press, 1987.

\bibitem{suhaimi2022investigating}
N.~M. Suhaimi, Y.~Zhang, M.~Joseph, M.~Kim, A.~G. Parker, and J.~Griffin.
\newblock Investigating older adults’ attitudes towards crisis informatics
  tools: Opportunities for enhancing community resilience during disasters.
\newblock In {\em Proceedings of the 2022 CHI Conference on Human Factors in
  Computing Systems}, CHI '22. Association for Computing Machinery, New York,
  NY, USA, 2022. doi: {{%
10\hspace{.1pt}\discretionary{.}{%
}{.}\hspace{.4pt}1145\discretionary{/}{%
}{/}3491102\hspace{.1pt}\discretionary{.}{%
}{.}\hspace{.4pt}3517528}}


\bibitem{thomas2006general}
D.~R. Thomas.
\newblock A general inductive approach for analyzing qualitative evaluation
  data.
\newblock {\em American journal of evaluation}, 27(2):237--246, 2006.

\bibitem{tory2021finding}
M.~Tory, L.~Bartram, B.~Fiore-Gartland, and A.~Crisan.
\newblock Finding their data voice: Practices and challenges of dashboard
  users.
\newblock {\em IEEE Computer Graphics and Applications}, 2021. doi: {{%
10\hspace{.1pt}\discretionary{.}{%
}{.}\hspace{.4pt}1109\discretionary{/}{%
}{/}MCG\hspace{.1pt}\discretionary{.}{%
}{.}\hspace{.4pt}2021\hspace{.1pt}\discretionary{.}{%
}{.}\hspace{.4pt}3136545}}


\bibitem{states_health_departments}
{USA Gov}.
\newblock State health departments.
\newblock https://www.usa.gov/state-health.

\bibitem{van2014communicate}
S.~L. Van~der Linden, A.~A. Leiserowitz, G.~D. Feinberg, and E.~W. Maibach.
\newblock How to communicate the scientific consensus on climate change: plain
  facts, pie charts or metaphors?
\newblock {\em Climatic Change}, 126(1):255--262, 2014. doi: {{%
10\hspace{.1pt}\discretionary{.}{%
}{.}\hspace{.4pt}1007\discretionary{/}{%
}{/}s10584\discretionary{%
}{-}{-}014\discretionary{%
}{-}{-}1190\discretionary{%
}{-}{-}4}}


\bibitem{van2021public}
L.~J. Van~Scoy, B.~Snyder, E.~L. Miller, O.~Toyobo, A.~Grewel, G.~Ha,
  S.~Gillespie, M.~Patel, J.~Reilly, A.~E. Zgierska, et~al.
\newblock Public anxiety and distrust due to perceived politicization and media
  sensationalism during early covid-19 media messaging.
\newblock {\em Journal of Communication in Healthcare}, 14(3):193--205, 2021.
  doi: {{%
10\hspace{.1pt}\discretionary{.}{%
}{.}\hspace{.4pt}1080\discretionary{/}{%
}{/}17538068\hspace{.1pt}\discretionary{.}{%
}{.}\hspace{.4pt}2021\hspace{.1pt}\discretionary{.}{%
}{.}\hspace{.4pt}1953934}}


\bibitem{walny2019data}
J.~Walny, C.~Frisson, M.~West, D.~Kosminsky, S.~Knudsen, S.~Carpendale, and
  W.~Willett.
\newblock Data changes everything: Challenges and opportunities in data
  visualization design handoff.
\newblock {\em IEEE Transactions on Visualization and Computer Graphics},
  26(1):12--22, 2020. doi: {{%
10\hspace{.1pt}\discretionary{.}{%
}{.}\hspace{.4pt}1109\discretionary{/}{%
}{/}TVCG\hspace{.1pt}\discretionary{.}{%
}{.}\hspace{.4pt}2019\hspace{.1pt}\discretionary{.}{%
}{.}\hspace{.4pt}2934538}}


\bibitem{waser2011nodes}
J.~Waser, H.~Ribicic, R.~Fuchs, C.~Hirsch, B.~Schindler, G.~Bloschl, and
  E.~Groller.
\newblock Nodes on ropes: A comprehensive data and control flow for steering
  ensemble simulations.
\newblock {\em IEEE Transactions on Visualization and Computer Graphics},
  17(12):1872--1881, 2011. doi: {{%
10\hspace{.1pt}\discretionary{.}{%
}{.}\hspace{.4pt}1109\discretionary{/}{%
}{/}TVCG\hspace{.1pt}\discretionary{.}{%
}{.}\hspace{.4pt}2011\hspace{.1pt}\discretionary{.}{%
}{.}\hspace{.4pt}225}}


\bibitem{armchair2020Weinman}
S.~Weinman.
\newblock The dangerous rise of covid-19 influencers and armchair
  epidemiologists.
\newblock
  https://www.insidehook.com/article/news-opinion/david-dunning-armchair-epidemiologists-coronavirus,
  2020.

\bibitem{welhausen2015visualizing}
C.~A. Welhausen.
\newblock Visualizing a non-pandemic: Considerations for communicating public
  health risks in intercultural contexts.
\newblock {\em Technical Communication}, 62(4):244--257, 2015.

\bibitem{winner1980artifacts}
L.~Winner.
\newblock Do artifacts have politics?
\newblock {\em Daedalus}, pp. 121--136, 1980.

\bibitem{zhang2020understanding}
Y.~Zhang, N.~Suhaimi, R.~Azghandi, M.~A. Joseph, M.~Kim, J.~Griffin, and A.~G.
  Parker.
\newblock {\em Understanding the Use of Crisis Informatics Technology among
  Older Adults}, p. 1–13.
\newblock Association for Computing Machinery, New York, NY, USA, 2020.

\bibitem{zhang2022shifting}
Y.~Zhang, N.~M. Suhaimi, N.~Yongsatianchot, J.~D. Gaggiano, M.~Kim, S.~A.
  Patel, Y.~Sun, S.~Marsella, J.~Griffin, and A.~G. Parker.
\newblock Shifting trust: Examining how trust and distrust emerge, transform,
  and collapse in covid-19 information seeking.
\newblock In {\em Proceedings of the 2022 CHI Conference on Human Factors in
  Computing Systems}. ACM, New York, NY, USA, 2022. doi: {{%
10\hspace{.1pt}\discretionary{.}{%
}{.}\hspace{.4pt}1145\discretionary{/}{%
}{/}3491102\hspace{.1pt}\discretionary{.}{%
}{.}\hspace{.4pt}3501889}}


\bibitem{zhang2021mapping}
Y.~Zhang, Y.~Sun, L.~Padilla, S.~Barua, E.~Bertini, and A.~G. Parker.
\newblock Mapping the landscape of covid-19 crisis visualizations.
\newblock In {\em Proceedings of the 2021 CHI Conference on Human Factors in
  Computing Systems}, CHI '21. Association for Computing Machinery, New York,
  NY, USA, 2021. doi: {{%
10\hspace{.1pt}\discretionary{.}{%
}{.}\hspace{.4pt}1145\discretionary{/}{%
}{/}3411764\hspace{.1pt}\discretionary{.}{%
}{.}\hspace{.4pt}3445381}}


\bibitem{zhao2022visualization}
L.~Zhao and W.~Ye.
\newblock Visualization as infrastructure: China’s data visualization
  politics during covid-19 and their implications for public health
  emergencies.
\newblock {\em Convergence}, p. 13548565211069872, 2022. doi: {{%
10\hspace{.1pt}\discretionary{.}{%
}{.}\hspace{.4pt}1177\discretionary{/}{%
}{/}13548565211069872}}


\end{thebibliography}
\end{document}